\documentstyle[12pt]{article}

\catcode`\@=11

\def\@maketitle{\newpage   % Pull out definition from article.sty
 \null
 \vspace*{-1\headsep}      % Move up to header region
 \vspace*{-1\headheight}
 \vspace*{-24pt}
 \begin{flushright}{\absize
   { \preprintno} \\ \@date}
 \end{flushright}
 \vskip \headsep	   % Move back down by part of what we moved up
 \vskip \headheight
 \begin{center}		   % Go back to original definition from article.sty
   {\tisize\bf \@title \par}
   \vskip 2em
   {\ausize
     \begin{tabular}[t]{c}\@author
     \end{tabular}\par}
   \vskip 1ex
 \end{center}
 \par
 \vskip 2ex}

\newcommand{\preprintno}{preprint number here}	 % Give a default preprint
\def\abstract{\if@twocolumn
\section*{Abstract}
\else \absize
\begin{center}
{\bf Abstract\vspace{0pt}}
\end{center}
\setskip
\quotation
\fi}
\def\endabstract{\if@twocolumn\else\endquotation\fi}

\def\section{
\setcounter{equation}{0} 	% Reset eqn numbers at start of section
\@startsection {section}{1}{\z@}{-3.5ex plus -1ex minus
 -.2ex}{2.3ex plus .2ex}{\Large\bf}}

\catcode`\@=12

\def\tisize{\Large}      %Title will be ambx12 at magstep1
\def\ausize{\large}      %Author names will be amr12 at magstephalf
\def\absize{\normalsize} %Abstract will be amr12
\def\setskip{ \setlength{\baselineskip}{4ex} } %Sets baselineskip in abstract

\def\starttext{
\setlength{\baselineskip}{ 17pt} %17pt for preprints, 25pt for Publications
\pagenumbering{arabic}
}

\let\Ga=\Gamma

\let\De=\Delta

\let\la=\lambda
\let\La=\Lambda

\let\del=\nabla

\let\th=\theta

\let\p=\partial
\let\<=\langle
\let\>=\rangle

\let\txt=\textstyle
\let\dsp=\displaystyle

\def\beq{\begin{equation}}
\def\eeq{\end{equation}}
\def\ba{\begin{array}}
\def\bea{\begin{eqnarray}}
\def\ea{\end{array}}
\def\eea{\end{eqnarray}}

\def\comment#1{ \hbox{[{\it Comment suppressed here.}\/]} }

\newcommand{\lapp}{ {\txt {{\txt <} \atop {\txt \sim}}} }
\newcommand{\gapp}{ {\txt {{\txt >} \atop {\txt \sim}}} }

\def\eqn#1{(\ref{#1})}  %Sticks parentheses around the LaTex "ref" macro
\def\Vdyn{V_{\rm dyn}}
\def\Fdyn{F_{\rm dyn}}
\def\half{{\txt{1\over 2}}}

\begin{document}

%Title page:

\title{Metastability in Two Dimensions and the Effective Potential}

\author{
        Mark Alford\\[1ex]
	Newman Lab of Nuclear Studies\\
	Cornell University\\
	Ithaca, NY 14853\\[3ex]
\and
	Marcelo Gleiser\\[1ex]
	Department of Physics and Astronomy\\
	Dartmouth College\\
	Hanover, NH 13755
}
\date{\today}

\renewcommand{\preprintno}{CLNS 93/1202\\ DART-HEP-93/03}

\begin{titlepage}

\maketitle

\def\thepage {}        % Kill page numbering for title page

\begin{abstract}

We study analytically and numerically the decay of
a metastable phase in (2+1)-dimensional classical
scalar field theory coupled to a
heat bath, which is equivalent to two-dimensional Euclidean
quantum field theory at zero temperature.
By a numerical simulation we obtain the
nucleation barrier as a function of the parameters of the potential,
and compare it to the theoretical prediction from the bounce (critical bubble)
calculation. We find the nucleation barrier to be accurately
predicted by theory using the bounce configuration obtained from the
tree-level (``classical'') effective action. Within the range of parameters
probed, we found that  using the bounce
derived from the one-loop effective action
requires an unnaturally large prefactor to match the lattice results.
Deviations from the tree-level prediction are seen in the
regime where loop corrections would be expected to become
important.

\end{abstract}

%PACS numbers: 11.10.Lm, 64.60.Qb, 05.40.+j, 98.80.Cq

\end{titlepage}

\starttext

\section {Introduction}

In the early eighteenth century, Gabriel Daniel Fahrenheit \cite{Fahren}
noticed that pure water could be cooled well below 32$^{\rm o}$
on his newly invented temperature scale, and still remain a liquid.
It was not until 1935, however, that
Becker and D\"oring \cite{BD}
gave a quantitative nucleation-theoretic
treatment of droplet formation for fluid systems.
Phenomenological field theory treatments were developed by Cahn and
Hilliard \cite{CH},
and by Langer \cite{Langer} within the context of a coarse-grained
Ginzburg-Landau model.
In quantum field theory, the study of metastable
vacuum decay was initiated by Voloshin, Kobzarev, and Okun \cite{Voloshin},
and put onto firm
theoretical ground by Coleman and Callan~\cite{Coleman}.
The realm of applicability of homogeneous
nucleation theory is extremely wide, from vapors, liquids, and solutions to
metals, polymers, and glasses \cite{general}.
Recently, the possibility that first-order phase
transitions occurred in the early Universe has generated a great
deal of interest in the metastable decay of the vacuum itself.
Well known examples are inflation \cite{inflation},
the electroweak phase transition \cite{eweak},
and the quark-hadron phase transition \cite{QH}.

Surprisingly, however, classical nucleation theory has yet to
receive clear experimental verification \cite{experiment}:
most experimental systems have complicated features that
are ignored when formulating a Ginzburg-Landau description
to which the classical theory could be applied.
Furthermore, even when a reliable field-theoretic
description is known, it is unclear how to accurately
obtain the coarse-grained free-energy functional that
is required for the critical-bubble calculation
\cite{Langer_MS}.
In cosmological applications, it has been customary to use
the one-loop effective potential in the calculation of the
nucleation barrier \cite{inflation,linde}.
However, this procedure has been questioned, since
$V_{\rm 1-loop}$ only reflects static properties of the theory,
and hence may not be a good guide to the dynamics of
out-of-equilibrium processes \cite{heretics}. Also, the incorporation of
temperature corrections to the nucleation barrier has not been carefully
considered \cite{GMR}.

Given the universality of the topic it would seem desirable to
test the classical theory within its area of validity
by performing a numerical simulation of some simple model,
for which we can measure the effective nucleation barrier. This barrier would
then be compared with theoretical predictions.
In recent work \cite{afg} we carried out this programme for a
one-dimensional Ginzburg-Landau model, and showed that classical nucleation
theory using the microscopic (lattice) Hamiltonian
agrees well with numerical simulations. In this letter we will
analyze the more realistic two-dimensional case, for which the lattice
Hamiltonian is cutoff-dependent, and hence it cannot be used to
determine the nucleation barrier.

The system we study is
classical thermal (2+1)-dimensional scalar field theory,
which is closely related to zero-temperature
quantum scalar field theory in two-dimensional Euclidean spacetime.
Replacing the temperature $T$ with the quantum of action $\hbar$,
the equilibrium correlation functions
of the classical theory map exactly onto the Green's
functions of the quantum theory, and the free energy maps onto
the effective action. Since we will find it convenient to mix the
classical and quantum terminologies, we give a dictionary
in Table 1.
Even though the classical nucleation rate
in Langevin time seems like a non-equilibrium quantity with
no quantum analogue, classical nucleation theory
identifies it (up to a dynamical factor) with the imaginary part
of the equilibrium free energy density of the
metastable state \cite{Langer}. Under $T\to\hbar$ this maps onto
the quantum tunnelling rate, which is given in terms of the
imaginary part of the energy density of the false vacuum \cite{Coleman}.
Therefore,
one could also think of our simulation as an independent way of
calculating the barrier to zero temperature quantum tunnelling.
In the classical language, we will show that the barrier obtained from
the tree-level free energy (the Hamiltonian), ignoring entropic (loop)
corrections, is in excellent agreement with the numerical results.
In the quantum language, what we will show is that the tree-level
(classical) action yields the correct bounce action in the
Callan and Coleman calculation.
Naturally this result may be modified when
the scalar field interacts with other fields \cite{erick,GMR}.

\begin{table}[tb]
% Struts to give correct row spacing in table:
\def\st{\rule[-2ex]{0em}{5ex}}
\def\bigst{\rule[-3.25ex]{0em}{7.5ex}}
\def\hugest{\rule[-7ex]{0em}{11ex}}
\let\scr=\scriptstyle

\begin{center}
\begin{tabular}{|c|c|}\hline
\st\bf Classical Statistical Mechanics & \bf Euclidean Quantum Field Theory\\
\hline
  \bigst
  $\ba{c} \hbox{Microscopic (Lattice) Hamiltonian} \\
     H_{\rm latt}= \int {\cal H}_{\rm latt}\,d^2x
  \ea $  &
  $\ba{c} \hbox{Bare Action} \\
     S_{\rm bare}= \int {\cal L}_{\rm bare}\,d^2x
  \ea $ \\
\hline
  \bigst
  $\ba{c} \hbox{Lattice Energy Density} \\  {\cal H}_{\rm latt} =
    \half Z_{\rm latt}|\del \phi|^2 + V_{\rm latt}(\phi;a)\ea $ &
  $\ba{c} \hbox{Bare Lagrangian Density}  \\  {\cal L}_{\rm bare} =
    \half Z_{\rm bare}|\del \phi|^2 + V_{\rm bare}(\phi;a)\ea $ \\
\hline
%\st Lattice Potential~ $V_{\rm latt}(\phi;a)$ &
%    Bare Potential~ $V_{\rm bare}(\phi;a)$ \\
%\hline
  \bigst
  $\ba{c} \hbox{Free Energy} \\
     F = \int {\cal F}\,d^2x
  \ea $  &
  $\ba{c} \hbox{Effective Action} \\
     S_{\rm eff}= \int {\cal L}_{\rm eff}\,d^2x
  \ea $ \\
\hline
  \bigst
  $\ba{c} \hbox{Free Energy Density} \\
     {\cal F} = \half Z_F|\del \phi|^2 + V_F(\phi)
  \ea $   &
  $\ba{c} \hbox{Effective Lagrangian Density} \\
     {\cal L}_{\rm eff} = \half Z_{\rm eff}|\del \phi|^2
                 + V_{\rm eff}(\phi)
  \ea $ \\
\hline
%\st $Z$ & Effective Field Normalization~ $Z_{\rm eff}$ \\
 $F(\phi,T) = \!\!
  \underbrace{\vphantom{\half}H(\phi)}_{\rm Hamiltonian} \!\! +
  \underbrace{\sum_{n=1}^\infty \De F_{n-{\rm loop}}(\phi)\,T^n}_{\rm
    Entropic~corrections}$
   &   $S_{\rm eff}(\phi,\hbar) =
  \underbrace{\vphantom{\half}S_{\rm class}(\phi)}_{
         \scr{\rm Classical}\atop\scr{\rm action}}+
  \underbrace{\sum_{n=1}^\infty \De S_{n-{\rm loop}}(\phi)\,\hbar^n}_{\rm
    Quantum~corrections}$\\[0ex]
  $V_F(\phi,T)= U(\phi) \,+\,
  \overbrace{\sum_{n=1}^\infty \De V_{n-{\rm loop}}(\phi)\,T^n}$
   &   ~$V_{\rm eff}(\phi,\hbar) =
  \underbrace{U(\phi)}_{\scr{\rm Classical}\atop\scr{\rm potential}} + \,
  \overbrace{\sum_{n=1}^\infty \De V_{n-{\rm loop}}(\phi)\,\hbar^n}$
\\[0ex]
\hline
\end{tabular}
\end{center}
\caption{ Dictionary of
equivalent terms for classical thermal and quantum mechanical
two-dimensional field theory on a lattice with spacing $a$. }
\end{table}

There has been only one previous numerical study of nucleation,
that of Valls and
Mazenko \cite{Maz}, which found that the theory gave a very poor
estimate of the nucleation barrier in a two-dimensional Ginzburg-Landau model
coupled to a heat bath.
There are several reasons \cite{afg} why our results should differ from
theirs, the dominant one being that
two-dimensional classical thermal field theory suffers from
ultra-violet divergences which produce severe lattice-spacing
dependence of many measured quantities, including nucleation rates.
Continuum results can be obtained, as we will show,
by introducing cutoff-dependent
counterterms and renormalizing. Since Valls and Mazenko did
not explicitly introduce counterterms, it is possible that
the continuum theory they were simulating was different from the
one for which they were calculating theoretical decay rates.

We will now review the ``critical bubble''
or ``bounce'' method of calculating the decay rate,
before proceeding to test it numerically.

\section{The Theoretical Results}

Consider a classical (2+1)-dimensional scalar field theory
with an asymmetric
double-well potential, at finite temperature.
By virtue of its relation to a two-dimensional quantum field theory,
mentioned above, we expect this system
to be formally ill-defined because of ultra-violet divergences in
the correlation functions. For the moment we therefore imagine integrating
out the high-momentum physics, and assume that the low-momentum dynamics
is described by some coarse-grained free-energy functional,
\beq
\dsp {1\over T} \Fdyn[\phi] =
  \dsp {1\over T} \int \Bigl\{ \half Z_{\rm dyn} |\del\phi|^2
  +  \Vdyn(\phi)
\Bigr\} \,d^2x~~.
\eeq
We assume that $\Vdyn(\phi)$
has a local minimum at $\phi=0$,
separated by a barrier from a lower global minimum at
some positive (perhaps infinite) $\phi$.

In general one does not know what $\Fdyn$ should be.
Langer \cite{Langer_MS} states that it should be a free
energy coarse-grained up to the correlation length
(but no further). We will investigate two candidates for
$\Fdyn$, the tree-level free energy or Hamiltonian $H$
(the classical action, in the quantum language)
and the 1-loop free energy (1-loop effective action) (see Table 1).

We will study nucleation for potentials of the general form
\beq
U(\phi)=
\half m^2\phi^2 - {\txt{1\over 6}} g \phi^3 + {\txt{1\over 24}} h \phi^4~~.
\eeq
This seems to give a 4-parameter family of theories, but in fact
by rescaling the field and distance, we can reduce it to a
functional involving only two (dimensionless) parameters, $\la$ and $\th$:
\beq\label{hamil}
\dsp {1\over T} H[\phi] =
\dsp {1\over \th} \int \Bigl\{ \half |\del\phi|^2
  +  \half\phi^2 - {\txt{1\over 6}} \phi^3
+ {\txt{1\over 24}} \la \phi^4 \Bigr\} \,d^2x~~,
\eeq
where
\beq
\th = {g^2 T \over m^4} \quad \hbox{and} \quad \la = {h m^2\over g^2}
\eeq
The dimensionless parameter $\th$ plays the role of temperature,
while $\la$ parameterizes the different potentials.
The potential has one minimum at $\phi=0$, and
for $0< \la < {1\over 3}$ there is a second, lower
minimum at positive $\phi$, separated from the metastable
minimum by a maximum
at $\phi_{\rm max}=(3/2\la)\Bigl(1-\sqrt{1-8\la/3}\Bigr)$.

For such a system, we expect $\Fdyn$ to be a functional
parameterized by $\la$ and $\th$, such that if the system starts out with
thermal expectation $\< \phi(\vec x)\>=0$ at finite temperature $\th$
then it will make a thermally activated transition to the
global minimum with a characteristic rate per unit area
\beq\label{rate}
\Ga = A \exp(-B/\th)~~,
\eeq
where $A$, the prefactor, is a function of $\la$, $\th$, and the
coupling to the heat bath $\eta$ (see Sect.~3). $B$ is the nucleation barrier,
which, according to the classical nucleation or bounce calculation
\cite{Langer,Coleman}, is the energy of the
``critical bubble'' or ``bounce'' saddle point of $\Fdyn$.
The critical bubble configuration $\phi_B(\vec x)$ is given by the static
(circularly symmetric) solution to the equation of motion,
\beq\label{eofm}
{\p^2\phi_B\over \p r^2} + {1\over r}{\p \phi_B\over \p r}
  =Z_{\rm dyn}^{-1}\,{\p \Vdyn\over \p\phi}(\phi_B)~~,
\eeq
with the boundary conditions that the field be
regular at the origin ($\phi_B'(0)=0$) and in the false vacuum
at spatial infinity ($\phi_B(\infty)=0$).
If one thinks
of $r$ as being ``time'', this equation describes the motion of a particle
with position $\phi_B$
in the potential $-\Vdyn(\phi)$, with a frictional resistance that is
inversely proportional to the time.
In general there is no closed-form solution to Eq.~\eqn{eofm},
but it can be solved numerically by guessing
a value for $\phi_B(0)$ and evolving forward in $r$. If the
solution overshoots and $\phi_B$ becomes negative
then the initial value was too
high; if $\p\phi_B / \p r$ becomes positive then it was too low.
In this way we converge on the correct solution, and its energy
is the nucleation barrier $B$.
Since in general $\Fdyn$ may depend on
$\th$, we may write $B=B_0 + B_1\th + B_2\th^2+\cdots$,
and so Eq.~\eqn{rate} becomes
\beq\label{genrate}
\Ga = A \exp(-B_1)\,\exp \left[-\left (B_0/\th+ B_2\th +
         \cdots\right)\right]~~.
\eeq

As was mentioned above, the tree-level candidate for $\Fdyn$ is
the Hamiltonian $H$.
This is independent of $\th$, so $B$ will just be $B_0$,
and we can calculate it by the method described above,
setting $\Vdyn=U$.
Another possibility is that $\Fdyn$ is $H$ plus some
entropic corrections. For example, in
studies of cosmological phase transitions it is generally assumed to
be the real part of the 1-loop free energy.
In this case $B_1$, $B_2$ etc are non-zero, but as
$\th\to 0$ the entropic corrections disappear, so $B_0$ is still
given by the tree-level ($F_{\rm dyn} = H$) barrier prediction.
The lowest order correction $B_1$
only modifies the prefactor, not the measured barrier.
Thus entropic corrections will primarily be visible through the effects
of the $B_2$ term, since we have little
information about the theoretical value of $A$ beyond the fact that
we expect it to be of order $1$ on dimensional grounds.
It will turn out that we can confirm that $B_0$ is given
by the tree-level bounce, but our observations cover too narrow a range
of $\th$ for the $B_2$ predicted by the 1-loop free energy to
be visible, so we are unable to say definitively whether
it gives the right entropic corrections.

We expect the barrier to go to
infinity when the two minima are degenerate, at $\la={1\over 3}$,
and to go to some finite value as $\la\to 0$.
In fact, $\la$ can take on negative values
and everything
still works exactly as described above. There is no longer a
global minimum to the potential, but there is still a metastable
state which decays by surmounting a barrier
that can be calculated by finding the critical bubble. We will
restrict ourselves to studying $\la$ in the range $0$ to $0.32$,
since negative $\la$ is not physically relevant, and as
$\la\to {1\over 3}$ the barrier becomes large, requiring
high temperatures to induce nucleation, which invalidates the
loop expansion we use in the next section.

\section{The Lattice Formulation}

In order to test the critical bubble theory, we need to simulate
the classical dynamics of the (2+1)-dimensional scalar field theory
in contact with a heat bath at temperature $\th$. This may be
done by evolution of the stochastic Langevin equation,
\beq\label{langevin}
{\p^2\phi\over\p t^2} = Z_{\rm latt} \del^2\phi - \eta {\p\phi\over\p t}
- {\p V_{\rm latt} \over \p \phi} + \xi(x,t)~~,
\eeq
where $\eta$ is the viscosity coefficient,
and $\xi$ is the stochastic noise with vanishing mean, related to
$\eta$ by the fluctuation-dissipation theorem,
\beq\label{flucdis}
\<\xi(\vec x,t)\xi( \vec x',t')\>=
  2\eta \th\delta(t-t')\delta^2(\vec x - \vec x')~~.
\eeq
In writing Eqns.~\eqn{langevin} and \eqn{flucdis}, we implicitly assumed
that the system is Markovian, {\it i.e.}, the correlation
time scale for the noise is much smaller than the typical relaxation time for
the system, which is the inverse of the
oscillation frequency around the metastable equilibrium.

In principle it now seems straightforward to put the field theory
on a two-dimensional lattice, and solve Eq.~\eqn{langevin}
numerically. However, \eqn{langevin} is expressed in terms
of the lattice (bare) parameters, so in order to make contact with the
calculation of Sect.~2 we must explicitly
construct a lattice potential
that has a definite continuum limit (lattice spacing $a\to 0$),
and have some way of specifying which continuum theory it
corresponds to. Since the theory has exactly the same divergences as
a two-dimensional Euclidean quantum field theory,
the lattice potential will have to contain lattice-spacing-%
dependent counterterms in order to give a good continuum limit.
(For a review, see Parisi \cite{Parisi}, Ch.~5).

Our strategy will be to
use continuum field theory with a hard momentum cutoff $\La\sim a^{-1}$
as our guide. We can easily calculate the free energy
(effective action) for
such a theory with a given bare potential by a loop expansion,
which is an expansion in powers of $\th$ ($\hbar$ in the quantum
theory). This will tell us what counterterms are needed to
cancel the cutoff dependence of the lattice theory.
We will only have to calculate to one loop, since there are
no divergent graphs beyond one loop.

We studied the lattice action defined by (see Table 1):
\beq\label{barepot}
\ba{rcl}
Z_{\rm latt} &=& 1~~, \\[1ex]
V_{\rm latt}(\phi) &=&  U(\phi) + V_{\rm ct}(\phi)~~,\\[1.25ex]
U(\phi) &=& \half\phi^2
- {\txt{1\over 6}} \phi^3 + {\txt {1\over 24}}\la \phi^4~~,\\[1.25ex]
V_{\rm ct}(\phi) &=& \dsp
 - {\th\over 4\pi}\ln\bigl(a M_1(\la,\th)\bigr) \phi
+ {\la\th\over 8\pi}\ln\bigl(a M_2(\la,\th)\bigr) \phi^2~~.
\ea
\eeq
By the standard calculation \cite{Sidney}
for a scalar quantum field theory in two Euclidean dimensions
with classical potential $U(\phi)$ and ultraviolet cutoff $\La$,
we find that the 1-loop effective potential is
\beq
V_F = U + V_{\rm ct}
  + {\th\over 2} \int_0^\La \!{d^2k \over (2\pi)^2}\,
\ln\biggl(1 + {U''\over k^2}\biggr)~~.
\eeq
Performing the integral and dropping terms independent of $\phi$,
we find for large $\La$,
\beq\label{effpot}
V_F = U + V_{\rm ct} +
  {\th\over 8\pi}\left( U''\ln(\La^2) + U'' - U''\ln(U'') \right)~~.
\eeq
The counterterms cancel the
$\La$-dependence, so we expect the theory
with lattice potential \eqn{barepot} to have
a good continuum limit. What effective potential will
this continuum theory have? Eq.~\eqn{effpot} was calculated for
a theory with momentum cutoff $\La$ to simulate the effects
of the lattice. It does not tell us what finite parts
will be left from the cancellation of the cutoff-dependence
in the actual lattice theory. Thus we can only say that we
expect the lattice theory defined by \eqn{barepot} to have
one-loop effective potential
\beq\label{oneloop}
V_F = \dsp U(\phi) + {\th\over 8\pi}
\biggl( U''(\phi) - U''(\phi) \ln\bigl(U''(\phi)\bigr) \biggr)
+ f(M_1)\phi + g(M_2)\phi^2~~.
\eeq
$M_1$ and $M_2$ are to be fixed by a renormalization condition,
but since the functions $f$ and $g$ are unknown, this will have
to be done numerically for the lattice theory. Our renormalization
condition was
\beq\label{rencond}
f(M_1) = g(M_2) = 0~~.
\eeq

It only remains to calculate the 1-loop effective field
normalization $Z$ (see Table 1). It is given in terms of
the one-particle-irreducible Green function $\Ga^{(2)}(p)$,
and is cutoff-independent in two dimensions.
Evaluating the relevant Feynman diagram,
\beq\label{equation}
\ba{rcl}
\dsp Z = {\p\Ga^{(2)}\over\p p^2}(0) &=&
\dsp 1 - {\th\over 48\pi} + {\cal O}(\th^2) \\[2ex]
\ea
\eeq

To summarize, the lattice theory \eqn{barepot} with
renormalization conditions \eqn{rencond} will have
one-loop free energy density (effective action)
\beq\label{free_en}
{\cal F} = \left(1-{\th\over 48\pi}\right)|\del\phi|^2 +
\dsp U(\phi) + {\th\over 8\pi}
\biggl( U''(\phi) - U''(\phi) \ln\bigl(U''(\phi)\bigr) \biggr)
\eeq
$U''$ is negative between the points of inflection of $U$, so ${\cal F}$
is complex. Weinberg and Wu \cite{ww} have suggested that the
real part of this is physically meaningful, so when using
${\cal F}$ in bounce calculations, we take the real part.

All these calculations are valid to one loop.
Without explicitly evaluating the higher corrections,
we can estimate that they will
be valid for $\th\ll 8\pi$, at least near the minima of
the potential.
There are no additional divergences at two
loops, but finite corrections to ${\cal F}$,
proportional to $\th^2$, will appear.
This puts a limit on how close we can push $\la$
to the value $1/3$, since the barrier becomes large as the
the two minima in $U(\phi)$ become degenerate, requiring
high temperatures in order to observe nucleations
on the lattice.

\section{The Numerical Analysis and Results}

We now have a way to test the theoretical calculation of Sect.~2.
For each value of $\la$ we measure the nucleation barrier
on the lattice by measuring
nucleation rates for a range of values of temperature $\th$.
For each $\la$ and $\th$ this involves two stages.
Firstly, we ensure that we are looking at the right continuum theory,
by running simulations with lattice potential \eqn{barepot},
and varying $M_1$ and $M_2$ until \eqn{rencond} is
obeyed. We then know that we have a theory whose free energy is given by
\eqn{free_en}, to first order in $\th/8\pi$.
The second stage involves measuring the
nucleation rate for this theory.
In both stages we ensure that we are studying the continuum
limit by reducing the lattice spacing
$a$ until the results become independent of $a$.
We checked the dependence of our results on the lattice
length, the time step,
the random number generator and the random number seed.
Within the limits of our numerical accuracy we found that the lattice
approximation correctly describes the continuum field theory.
We chose $\eta=1$ as the viscosity in all simulations.
In principle, knowledge of the physical nature of the heat bath
would enable one to calculate the value of $\eta$, as well as
additional nonlinear or nonlocal dissipative terms in the
Langevin equation.

The easiest way to impose the renormalization conditions
was to choose $M_1$ and $M_2$ such that the measured
true and false vacua on the lattice coincided with the
minima of the free energy \eqn{free_en}. This required
running a simulation of \eqn{langevin} on a relatively
small lattice ($L=10$), with initial conditions that
caused the field to settle down in the appropriate vacuum state.
It was then a simple matter to measure the thermal expectation
value of the field.
We found that the renormalization condition
$f(M_1)=0$ ({\it i.e.}~$\<\phi\>=0$ in the metastable state)
was obeyed for $M_1= 1/(2\pi)$, in the sense that with this choice
the measured metastable average field was always much less than
the peak $\phi_{\rm max}$.
The condition $g(M_2)=0$ was obeyed for
$M_2=1/(2\pi)$, in the sense that
the measured true vacuum was within a few
percent of the value predicted by \eqn{free_en}.
Obviously this method could not fix $M_2$ in the
$\la=0$ theory, since the true vacuum is then at infinite $\phi$.
However in that case there is no cutoff-dependent $\phi^2$ term
in the free energy, and $g(M_2)=0$.

To measure the nucleation rates, we evolved
the Langevin equation on square lattices of
size $L=20, 40, 60$ or 120, with lattice spacing $a=0.5$, using a leapfrog
algorithm with time step of $0.05$.
For each $\la$ and $\th$ we performed several hundred simulations.
In each one we started the system in the metastable state,
evolved the Langevin equation \eqn{langevin} forward in time,
measuring the time $t$ that elapsed before
it escaped from the metastable region
by nucleating a growing bubble of the stable state. From this data we
constructed a frequency histogram for nucleation times from which
$\Ga(\la,\th)$, the nucleation rate
per unit area, could be read off by fitting to the expected form:
${\rm Prob}(t)\propto \exp(-t\Ga L^2)$
for a lattice of size $L$. At low temperatures $\Ga$ gets small
\eqn{rate},
but we could compensate for this by increasing $L$. Thus with a range
of lattice sizes we were able to observe a wider range of
temperatures than would have been possible using only one lattice.
We were still constrained to a fairly narrow temperature range,
since the simulation ran much more slowly on larger lattices.

In order to determine when nucleation had occurred, we
calculated a smoothed field
by averaging $\phi$ over square blocks of size $\De L$.
The system was considered to have escaped from metastability
when one or more of the blocks achieved an average field value
greater than the peak $\phi_{\rm max}$ (see Sect.~2).
For each value of $\la$ we chose the smallest $\De L$
such that when one block was converted, the rest of the lattice
would always follow. For $\la=0, 0.1, 0.2$ we set $\De L = 5$;
For $\la=0.25, 0.3, 0.31$ we set $\De L = 10$; for $\la=0.32$ we set
$\De L = 20$.

Our results are summarized in Fig.~1, 2, and 3.

In Fig.~1 we show a logarithmic plot of
the inverse nucleation rate per unit area
$\Ga^{-1}$, as a function of inverse temperature, for
different values of the potential parameter $\la$. From
Eq.~\eqn{genrate} for
the decay rate we expect that $\ln(\Ga^{-1}) = \ln(A^{-1})+B_1
+ B_0(\la)/\th + B_2\th$.
The excellent straight line fits for $\la\leq 0.3$ indicate that
the classical theory that gave rise to Eq.~\eqn{genrate} is qualitatively
correct in this regime. There is no sign of a $B_2 \th$ term.
We conclude that the nucleation barrier is well approximated by the form
$B=B_0 + B_1\th$ for the ranges of temperature probed for $\la\leq 0.3$.
For $\la=0.31,\,0.32$ there are signs of a $B_2$ term, so
entropic corrections to the barrier are becoming important, as
one would expect at high $\th$.

In Fig.~2 we confront the numerical barrier measurements from Fig.~1
with the theoretical prediction for the nucleation barrier
$B_0$ given in Sect.~2, which involves taking $\Vdyn(\phi)$ to be $U(\phi)$,
the potential appearing in the tree-level free energy (Hamiltonian,
or classical action).  For $\la\leq 0.3$ we
obtain almost perfect quantitative agreement between theory and numerical
experiment, confirming that $B_0$ is accurately given by the action of
the tree-level bounce.
As one might expect from the weakness of their straight line fits in
Fig.~1, the barriers for $\la=0.31,\,0.32$ are not well predicted
by the tree-level bounce.

In Fig.~3 we plot the values of the prefactor $A$ that would fit
the data of Fig.~1, using the tree-level bounce and
the 1-loop bounce. For the tree-level points, we
fitted Eq.~\eqn{genrate} to the data of Fig.~1 with $B_1=0$,
and plotted the value of $A^{-1}$ for each value of $\la$.
For the 1-loop points we calculated
$B(\la,\th)$ using the method of Sect.~2 with $F_{\rm dyn}$ set to the
real part of the 1-loop free energy \eqn{free_en}.
(This was only possible for $\th \lapp 10$. For
larger $\th$ no solution could be found, so 1-loop barrier predictions
could not be given for $\la=0.31,\,0.32$, whose rate measurements
were all at larger $\th$.)
By fitting to the form $B=B_0(\la) + B_1(\la)\th$
we obtained $B_0$ (which agreed with the tree-level prediction,
as expected) and $B_1$. For each $\la$ we then found $A$ by
fitting Eq.~\eqn{genrate} to the data of Fig.~1.

Much larger prefactors were required to fit the 1-loop
predictions to the data than for the tree-level predictions.
Since $A$ is expected to be of order $1$, this provides
circumstantial evidence that the one-loop bounce is not
the relevant quantity. For a definitive answer, a more
powerful computer would be needed, allowing a larger range
of $\th$ to be probed for each $\la$, so that the value of
$B_2$ could be measured.

The fact that the results fit the theory less well as $\la$
approaches ${1\over 3}$ should not be surprising. Since the
barrier is diverging in this region, we are forced to
use high $\th$ in order to see any nucleations, and so we
expect entropic corrections to the barrier to become important.
The deviations begin at $\la=0.31$, where temperatures
in the range $12$ to $20$ are investigated.
This is approaching the regime $\th\gapp 8\pi$, where
the loop approximation breaks down, so we expect significant
$\th$-dependent corrections to the theoretical nucleation barrier.
In other words, we expect $B_2$ and higher terms to become large.
It is perhaps more surprising that we do not see deviations for
$\la=0.3$, where a similar temperature range was explored.

\section{Conclusions}

We have shown that, for a single scalar field in two space dimensions,
classical nucleation theory correctly predicts the nucleation
barrier to within a few percent, when the Hamiltonian,
{\it i.e.}~the tree-level free energy,
is used in the calculation. Equivalently (see Table 1), the
quantum bounce calculation correctly predicts the barrier, as long
as the tree-level (``classical'') action is used.

Obviously this statement only makes
sense for the parameter range where a loop expansion is possible,
and, as expected on the basis of the calculations of Sect.~3,
we found that it broke down for large loop parameter,
$\th/8\pi\gapp 1$. The interesting thing is that the tree-level action
gave correct predictions for $\th/8\pi$ as large as $0.5$,
a regime where one might have expected significant loop
corrections.

We were unable to make any definitive statement about the
1-loop bounce (the bounce solution obtained
by using the 1-loop effective potential). Fig.~3 shows that
it requires an unnaturally large prefactor, but the
temperature range we probed was too narrow to see if its
predicted higher-order corrections were present.
With the aid of more powerful computing resources it
should be possible to settle this question.

There are many other directions in which this work can be extended.
E.~Weinberg \cite{erick} has recently argued that a truncated 1-loop
effective potential correctly describes nucleation in
certain radiative symmetry breaking models, and it would
be interesting to study this numerically.
Another issue is finding
the dependence of the prefactor not only on the temperature but
also on the coupling to the heat bath. We have previously studied the
dependence of the thermal kink-antikink pair creation on the coupling to
the bath \cite{oldafg}, finding good agreement with Kramer's celebrated
result \cite{Kramers}. Perhaps of more relevance to current work on
out-of-equilibrium processes in the early Universe, we could study the
dynamics of weakly
first-order transitions. Recent work has claimed that for sufficiently weak
transitions the usual bubble nucleation mechanism discussed here must
be modified. Instead, it has been suggested that the transition may
evolve through the nucleation and subsequent percolation of sub-critical
bubbles in such a way as to resemble an emulsion of phases as the system
is cooled below the critical temperature \cite{GK}.
We are presently
investigating these questions.

\bigskip
\centerline{\bf Acknowledgments}
MG was supported in part by a National Science Foundation grant
No.~PHYS-9204726.
We are grateful to Mark Goulian, Farid Abraham, Ron Horgan, and
Peter Lepage for illuminating discussions.
MGA would like to thank the Materials Science Center
at Cornell for access to their network of IBM RS6000
workstations.

\bigskip
\centerline{\Large\bf Figure Captions}
\bigskip
\noindent
Figure 1: Numerically measured inverse nucleation rate $\Ga^{-1}$
as a function of inverse temperature,
with straight line fits, for a range of values of the dimensionless
$\phi^4$ coupling $\la$. From the right: $\la= 0,\, 0.1,\, 0.2,\,
0.25,\, 0.3,\, 0.31,\, 0.32$.

\bigskip
\noindent
Figure 2: Nucleation barrier as a function of $\la$.
The points are the slopes of the lines in Fig.~1. The
line is the critical bubble calculation of Sect.~2 at
tree level, {\em i.e.}~with $Z_{\rm dyn}=1$ and $\Vdyn=U$.

\bigskip
\noindent
Figure 3: Rate prefactor $A^{-1}$ as a function of $\la$.
Bottom set of points is for the tree-level
prediction (see Eq.~\eqn{rate}) fitted to the data of Fig.~1.
Top set of points is
for the 1-loop prediction (see Eq.~\eqn{genrate}).
We excluded $\la=0.31,\,0.32$ because their error bars were too large.

\end{document}